\def\cm{cm$^{-1}$}
\def\be{\begin{equation}}
\def\ee{\end{equation}}

\documentclass[twocolumn,prb,showpacs,amsmath,amssymb]{revtex4}
\usepackage{graphicx}
\usepackage{dcolumn}
\usepackage{bm}

\begin{document}

\title{Electronic properties of correlated metals in the vicinity of a charge order transition:\\
optical spectroscopy of $\alpha$-(BEDT-TTF)$_2M$Hg(SCN)$_4$ ($M$ =
NH$_4$, Rb, Tl)}
\author{N. Drichko$^{1,2}$}\author{M. Dressel$^1$}
\author{C. A. Kuntscher$^{1}$ $^*$} \author{~A. Pashkin$^{1}$}
\altaffiliation{Present address: Experimentalphysik\ II,
Universit\"at Augsburg, D-86135 Augsburg, Germany}
\author{A. Greco$^3$}
\author{J. Merino$^4$}
\author{J. Schlueter$^5$}
\affiliation{
$^1$ 1.~Physikalisches Institut, Universit{\"a}t Stuttgart, Pfaffenwaldring 57, 70550 Stuttgart, Germany \\
$^2$ Ioffe Physico-Technical Institute, St.Petersburg, Russia \\
$^3$ Facultad de Ciencias Exactas Ingenier\'ia y Agrimensura e Instituto de F\'isica Rosario (UNR-CONICET), Rosario, Argentina\\
$^4$ Dept. de Fisica Teorica de la Materia Condensada, Universidad
Autonoma de Madrid, Spain\\
$^5$ Material Science Division, Argonne National Laboratory,
Argonne, Illinois 60439-4831, U.S.A.}

\date{\today}

\begin{abstract}
The infrared spectra of the quasi-two-dimensional organic
conductors $\alpha$-(BEDT-TTF)$_2$\-$M$Hg(SCN)$_4$ ($M$ = NH$_4$,
Rb, Tl) were measured in the range from 50 to 7000~\cm\ down to
low temperatures in order to explore the influence of electronic
correlations in quarter-filled metals. The interpretation of
electronic spectra was confirmed by measurements of pressure
dependent reflectance of $\alpha$-(BEDT-TTF)$_2$\-KHg(SCN)$_4$ at
T=300~K. The signatures of charge order fluctuations become more
pronounced when going from the NH$_4$ salt to Rb and further to Tl
compounds. On reducing the temperature, the metallic character of
the optical response in the NH$_4$ and Rb salts increases, and the
effective mass diminishes. For the Tl compound, clear signatures
of charge order are found albeit the metallic properties still
dominate. From the temperature dependence of the electronic
scattering rate the crossover temperature is estimated below which
the coherent charge-carriers response sets in. The observations
are in excellent agreement with recent theoretical predictions for
a quarter-filled metallic system close to charge order.

\end{abstract}

\pacs{
71.10.Hf, 
71.30.+h, 
74.25.Gz, 
74.70.Kn  
}

\maketitle

\section{Introduction}

Electron-electron interactions are well recognized to be decisive
for the ground states observed in low-dimensional electronic
systems, for instance Mott insulator, charge-ordered state,
superconductivity, ferro- and antiferromagnetic order. The generic
phase diagram of correlated materials shows a number of important
features: the variation of crucial (order) parameters like the
electron doping or the effective electronic interactions drives
the system from a metal to an ordered state, with
superconductivity found around some critical
point.\cite{Mathur,Kotliar04} While in the high-temperature
superconductors the control parameter typically is the hole
doping, the quasi-two dimensional-organic conductors give a
possibility to nicely tune the effective electronic
interactions.\cite{Seo04} The well-developed  phase diagram of
half-filled BEDT-TTF-based organic conductors demonstrates this
approach: the ground state of systems changes from a metal (Fermi
liquid) to a Mott insulator on the increase of electronic
correlations, with superconductivity found in
between.\cite{Lefebvre,McKenzie97}

Besides the on-site Coulomb interaction $U$, for the
quarter-filled systems the nearest-neighbor electronic repulsion
$V$ is the second important parameter. In the case of large values
of $U$, a change of the ground state from metallic to a
charge-ordered on the increase of $V$ was proposed for the
quarter-filled layered molecular conductors; around the quantum
critical point superconductivity mediated by charge fluctuations
was suggested.\cite{Merino01} The family of quarter-filled organic
conductors $\alpha$-(BEDT-TTF)$_2M$Hg(SCN)$_4$ ($M$=NH$_4$, Rb,
Tl, K) is a prime candidate to experimentally study the vicinity
of a phase border between metallic and charge ordered states in
the phase diagram calculated in Ref.~\onlinecite{Merino01} and
schematically shown in Fig.~\ref{phasediagram} of this work: the
NH$_4$ compound exhibits superconductivity at $T_c\approx 1$~K,
while the others are metallic with a density-wave state observed
below 10~K, which is still subject to
discussion.\cite{Wosnitza96,Singleton00} In the present
investigation we are interested in the temperature region above
the density wave state where the band structure and Fermi surface
of these isostructural compound are basically identical.

These crystals are composed by alternating layers of two types:
conductivity occurs in layers of BEDT-TTF
[bis-(ethyl\-ene\-di\-thio)\-te\-tra\-thia\-ful\-va\-lene]
molecules, while the layers of the polymeric anions
[$M$Hg(SCN)$_4]^-$ serve as a `charge reservoir'; in the following
we refer to the various salts by their metal ions $M$=NH$_4$, Rb,
Tl, or K in the anion layer. They have the same valence but
different volume, which effects the size of the unit cell and
hence the transfer integrals $t$ between the BEDT-TTF molecules in
the conducting layer in some non-trivial way,\cite{Mori_Term}
leading to a different $V/t$ parameter between the compounds.

 A change of the control parameter $V/t$ by chemical modifications in the anionic layer was extensively
utilized by H. Mori {\it et al.}\cite{Mori98} when synthesizing
the quarter-filled family $\theta$-(BEDT-TTF)$_2MM'$(SCN)$_4$.
These quarter-filled compounds are closer to an insulating charge
order state, and the transition temperature between the metallic and
insulating phase decreases as the ratio $V/t$ shrinks.
A phase diagram was proposed based on resistivity studies
which exhibits the general features as mentioned
above.\cite{Mathur,Merino01} A number of experiments prove that
the charge disproportionation already develops in the  metallic
phase at temperatures well above the phase
transition.\cite{Chiba04,Suzuki03} In
$\theta$-(BEDT-TTF)$_2$RbZn(SCN)$_4$, for example, the width of
the lines observed in NMR experiments suggests a slowly
fluctuating charge order.\cite{Chiba04} On the other hand, the
only charge-order insulator of the $\alpha$-phase family,
$\alpha$-(BEDT-TTF)$_2$I$_3$,\cite{Schweitzer87,Dressel94}
 exhibits a timely stable charge disproportionation\cite{Moroto04} already
at temperatures above the insulating state.

While those investigations lowered the temperature in order to
move towards the metal-insulator transition, a particulary
interesting possibility would be to tune the $V/t$ ratio to
approach the phase transition to the insulating state from the
metallic side. Do charge-order fluctuations develop? Or do stable
regions form which are partly charge ordered? Is  this a first
order or a second order phase transition? Will charge-order
fluctuations be enhanced although the system remains metallic? The
$\alpha$-(BEDT-TTF)$_2M$Hg(SCN)$_4$ family is the proper system to
address these questions.

Optical investigations in a wide frequency range are most suitable
to characterize the charge dynamics and to identify deviation from
a simple metallic behavior.\cite{DresselGruner02,Basov05,Imada}
For quarter-filled systems close to charge order, the
frequency-dependent conductivity was calculated by  exact
diagonalization for large $U$ and different $V/t$
ratios.\cite{Merino03} The spectral weight is expected to shift
from the Drude peak to higher frequencies  as correlations
intensify. It is further proposed that the effective mass of the
charge carriers and their scattering rate depends on correlations
and temperature\cite{Merino03}. In previous optical studies clear
signatures of charge-order fluctuations have been observed in
$\alpha$-(BEDT-TTF)$_2$KHg(SCN)$_4$ in contrast to the
superconducting analog
$\alpha$-(BEDT-TTF)$_2$NH$_4$KHg(SCN)$_4$.\cite{Dressel03} It is
now of great interest to extend our investigations to the whole
family of $\alpha$-(BEDT-TTF)$_2M$Hg(SCN)$_4$ materials and to
compare the experimental findings with the theory. This gives us
insight on how quarter-filled two-dimensional metallic system
behave close to a correlation-driven charge-order transition.

The paper is structured in the following way, in Section~II we
present the experimental techniques, in Section~III we present the
results, a primary analysis of the observed spectra (B) and its
interpretation (C), and in (D) we show that pressure-dependent
measurements confirm our interpretation, in E and F we discuss the
nature of the major anisotropy and temperature-dependent effects.
In Section~IV we further interpret our results in terms of
metallic quarter-filled system close to charge order and an
ordered system. Our findings are summarized in Section~V
(conclusions.)

\section{Experimental Techniques}

Single crystals of $\alpha$-(BEDT-TTF)$_2M$Hg(SCN)$_4$ (M=NH$_4$,
Rb, Tl) were grown according to Ref.~\onlinecite{Mori90} and reach
up to $2\times2$~mm$^2$ in the highly conducting $ac$-plane; the
crystal structure was confirmed by X-ray diffraction. When
characterized by dc resistivity, the results coincide with
previous reports. The polarized reflectivity of the crystals was
measured in the conducting plane along the main optical axes,
parallel and perpendicular to the stacks of BEDT-TTF molecules
(i.e. parallel $c$ and $a$ crystal axes) in the frequency range
between 50 and 7000~\cm. The main axes were identified at room
temperature by polarization dependence measurements with an
accuracy of 2$^\circ$. The spectral resolution used  was 1~\cm\
for the NH$_4$ and Tl salts and 2~\cm\ in the case of
$\alpha$-(BEDT-TTF)$_2$RbHg(SCN)$_4$. The sample was cooled in a
cold-finger cryostat with a rate of circa 1~K/min; spectra were
taken at 300, 200, 150, 100, 50 and 6~K. In order to receive the
absolute values of reflectivity, the sample covered {\it in situ}
with 100~nm gold was used as a reference; this technique is
described in Ref.~\onlinecite{Homes93}. Although these
measurements agree with previously published
spectra\cite{Dressel92,Dressel03,Tajima03} as far as the overall
shape is concerned, the absolute values are up to 5\%\ higher due
to our superior method of measuring the absolute reflectivity
value. The mid-infrared data were also double checked at room
temperature using an infrared microscope with a spot size of
100~$\mu$m. We were not able to use this improved technique for
the low-temperature measurements of
$\alpha$-(BEDT-TTF)$_2$KHg(SCN)$_4$ due to poor quality and small
size of the crystals; thus we do not extend the present
quantitative analysis of the optical data to this compound.

The optical conductivity was evaluated by the Kramers-Kronig
analysis of the reflection spectra. The spectra in the 9000-40000
\cm\ were measured at room temperature using a home-made
microspectroreflectometer; at frequencies above 40000 \cm\ the
common $\omega^{-2}$ and $\omega^{-4}$ extrapolations were used,
while a Hagen-Rubens assumption was applied at low frequencies.
The agreement obtained with the dc conductivity is excellent. In
the case of the NH$_4$ salt, for instance, the measured dc values
range from $\sigma_{\rm dc}=100$ to $400~(\Omega{\rm cm})^{-1}$ at
ambient temperature and 100 times higher at $T=4.2$~K; the value
for optical conductivity at $\omega\rightarrow 0$ is about
$350~(\Omega{\rm cm})^{-1}$ at 300~K and $36000~(\Omega{\rm
cm})^{-1}$ at $T=6$~K.

Polarized reflectance measurements of
$\alpha$-(BEDT-TTF)$_2$KHg(SCN)$_4$ under pressure were performed
using a diamond anvil cell equipped with type IIA diamonds
suitable for infrared measurements. Finely ground CsI powder was
chosen as quasi-hydrostatic pressure medium. For the pressure
experiment a small piece (about 80 $\mu$m $\times$ 100 $\mu$m in
size)
was cut from a single crystal and placed in the hole 
of a steel gasket. A ruby chip was added for determining the
pressure by the ruby luminescence method.\cite{Mao86} The
pressure-dependent reflectance was studied in the mid-infrared
frequency range ($550-8000$~\cm) at room temperature using a
Bruker IFS 66v/S FT-IR spectrometer in combination with an
infrared microscope (Bruker IRscope II). Reflectance spectra were
measured at the interface between sample and diamond anvil (the
measurement geometry is illustrated in
Ref.~\onlinecite{Kuntscher05}); spectra taken at the inner
diamond-air interface of the empty cell served as the reference
for normalization of the sample spectra. The pressure-dependent
reflectivity spectra reported below refer to the absolute
reflectivity at the sample-diamond interface, calculated according
to $R_{s-d}(\omega)=R_{\rm dia}\times
I_{s}(\omega)/I_{d}(\omega)$, where $I_s(\omega)$ denotes the
intensity spectrum reflected from the sample-diamond interface and
$I_d(\omega)$ the reference spectrum of the diamond-air interface.
$R_{\rm dia}$ was calculated from the refractive index of diamond
$n_{\rm dia}$ to 0.167. For quantitative description of the
spectra, a Drude-Lorentz fit was performed. To get reliable fit
parameters, we simultaneously fitted the zero-pressure
reflectivity spectra taken inside the cell (with an diamond-sample
interface) and the spectra taken outside the cell with an
air-sample interface. The change of the parameters of the Drude
part and the Lorentz oscillators were then followed as a function
of pressure.

\section{Results and Analysis}

\subsection{Experimental results}

\begin{figure*}
\includegraphics[width=15cm]{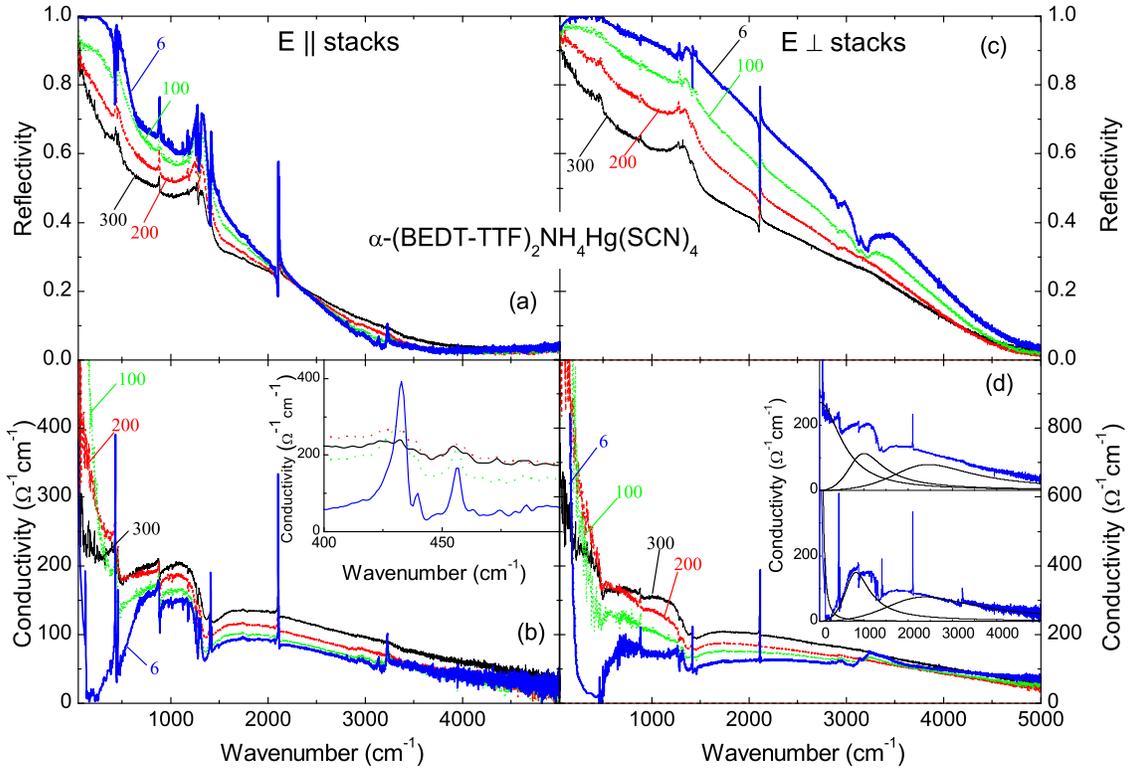}
\caption{Color online. Reflectivity and conductivity spectra of
$\alpha$-(BEDT-TTF)$_2$NH$_4$Hg(SCN)$_4$ at $T=300$, 200, 100 and
6~K; left panels (a) and (b): $E\parallel$~stacks, right panels
(c) and (d): $E\perp$~stacks. The inserts show optical
conductivity in the range of $\nu_{12}$(A$_g$) and
$\nu_{13}$(A$_g$) modes. The insert in (d) shows the contributions
of one Drude and two Lorentz components for $E\parallel$~stacks
300 K and 6 K. }\label{fig:NH}
\end{figure*}

\begin{figure*}
\includegraphics[width=15cm]{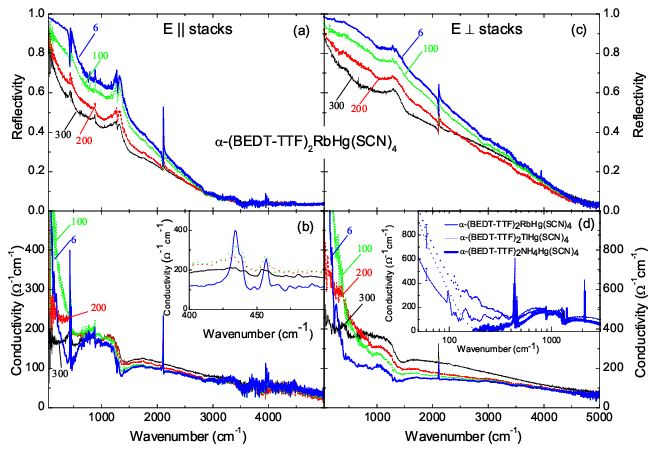}
\caption{\label{fig:Rb} Color online. Reflectivity and
conductivity spectra of $\alpha$-(BEDT-TTF)$_2$RbHg(SCN)$_4$ at
$T=300$, 200, 100 and 6~K; left frames (a) and (b):
$E\perp$~stacks, right frames (c) and (d): $E\parallel$~stacks.
The insert on (b) show optical conductivity in the range of
$\nu_{12}$(A$_g$) and $\nu_{13}$(A$_g$) modes. The insert in (d)
present the optical conductivity of the three salts in $E
\parallel$~stacks at 6 K in log scale. The low-frequency part
presented in dashed lines was received after smoothing the
reflectivity data, a respective error bar at 60 \cm\ is shown.}
\end{figure*}

\begin{figure*}
\includegraphics[width=15cm]{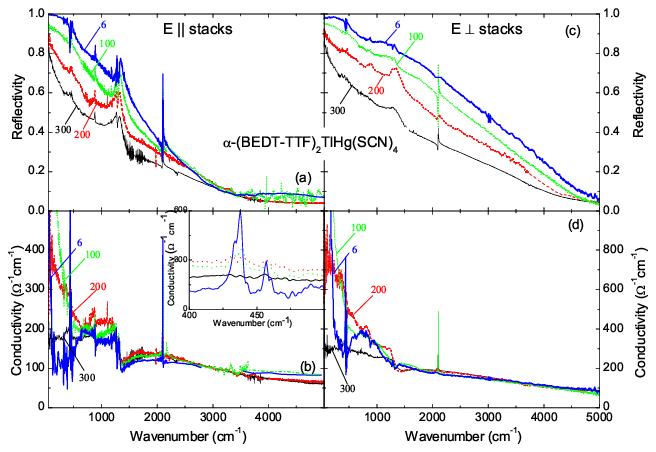}
\caption{ \label{fig:Tl} Color online. Reflectivity and
conductivity spectra of $\alpha$-(BEDT-TTF)$_2$\-TlHg(SCN)$_4$ at
$T=300$, 200, 100 and 6~K; panels (a) and (b): $E\perp$~stacks,
panels (c) and (d): $E\parallel$~stacks. The inserts show optical
conductivity in the range of $\nu_{12}$(A$_g$) and
$\nu_{13}$(A$_g$) modes.}
\end{figure*}

In Figs.~\ref{fig:NH} to \ref{fig:Tl} the optical reflectivity and
conductivity of the three compounds $\alpha$-(BEDT-TTF)$_2M$Hg(SCN)$_4$ ($M$=NH$_4$,
Rb, Tl) are displayed. Results for the K-analog have already been
presented in Ref.~\onlinecite{Dressel03}. The reflectivity of all
the compounds shows a metal-like frequency and temperature
behavior: a plasma edge is observed in the mid-infrared, and the
high reflectance at frequencies below the plasma edge increases
even further upon cooling. The higher reflectivity and
conductivity observed in the polarization $E\perp$~stacks is in
agreement with the calculated anisotropy of the transfer
integrals.\cite{Mori_Term} The room-temperature spectra of the Rb
and Tl salts differ only by few percent, while the NH$_4$ compound
shows a slightly higher reflectivity at low frequencies.

A number of vibrational features in the $400-1600$~\cm\ frequency
range are known to originate from the coupling of totally
symmetric A$_g$ vibrations of BEDT-TTF with electrons (emv
coupling).\cite{Drichko03} These comparatively weak emv-coupled
features evidence for the broken symmetry (dimerization) in the
stacks, in agreement with the X-ray structural data. The most
prominent of the emv-coupled features is the band of $\nu_4(A_g)$
vibration, which appears as a wide maximum between approximately
1200 and 1400~\cm. This feature is the same in all the salts,
while the lower frequency features show some differences between
the compounds, which will be discussed later in this paper. The
charge-transfer within the dimers, which activates the
totally-symmetric vibrations, is expected in the mid-infrared
region, but its intensity might be low, in agreement with the weak
vibrational features in comparison to the $\kappa$-phases, for
example.\cite{Kornelsen91,Dressel04,Faltermeier05} The narrow band
at about 2100~\cm\ is the infrared-active C-N stretching vibration
of the SCN groups in the anion layer.

At temperatures below 200~K the slightly different levels and
shapes of the far-infrared reflectance observed for the NH$_4$, Rb
and Tl salts lead to distinct differences in the conductivity
spectra. The low-temperature reflectivity of the NH$_4$ salt is
close to 100\%\ in the low-frequency limit but abruptly drops
above 700~\cm, especially apparent in the $E\parallel$~stack
polarization. This behavior converts to an intense and extremely
narrow zero-frequency peak in the conductivity spectrum; in
addition there are two wide bands at approximately 1000~\cm\ and
2500~\cm. Accordingly, the lower reflectivity level and gentler
slope obtained for the Rb salt causes a less pronounced and
wider zero-frequency peak and more spectral weight in the
high-frequency features. The reflectivity of the Tl salt also
approaches 100\%, but in a more gradual fashion; a bump at about
1000~\cm\ is superimposed on the high reflectance background at
low temperatures. Accordingly, in the conductivity spectra the
Drude-peak is not extremely narrow, and the 1000~\cm\ maximum is
more pronounced than in the spectra of the other two salts. This
tendency is even enhanced in the spectra of
$\alpha$-(BEDT-TTF)$_2$KHg(SCN)$_4$, as described in
Ref.~\onlinecite{Dressel03}.

\subsection{Drude-Lorentz analysis}
\label{DrudeLorentz} We use a Drude-Lorentz fit to disentangle
different contributions to the spectra; simultaneous fits of the
reflectivity and conductivity spectra put additional restrictions
on the parameters. An insert
 in Fig.~\ref{fig:NH}(d) is an example of the fit,
 showing the main electronic features in the
conductivity spectra of $\alpha$-(BEDT-TTF)$_2$NH$_4$Hg(SCN)$_4$
at $T=300$ and 6~K. The zero-frequency peak is approximated by a
simple Drude contribution; the maximum around 1000~\cm\ and the
wide band in the mid-infrared range are fitted by  a Lorentzian
each, yielding the central frequencies $\omega_t$ of 1000 and 2500
\cm, respectively. Only when the Drude-like feature sharpens below
200~K, the 1000~\cm\ contribution turns into a real peak. The
assessment of the Drude plasma frequency by zero-crossing of
$\epsilon_1$\cite{DresselGruner02} in agreement with the fit,
however, clearly reveals that it is already present at $T=300$~K;
in contrast to the case of half-filled BEDT-TTF-based compounds
where it develops only for $T<50$~K.\cite{Dressel04,Faltermeier05}

The simple Drude formula does not describe perfectly the shape of
the zero-frequency  conductivity peak, but gives reliable values
of the scattering rate (as a half-width of the peak). Since our
measurements go down to only 50~\cm, and the zero-frequency
conductivity peak becomes very narrow at low temperatures, we did
not carry out an extended Drude analysis,\cite{DresselGruner02} to
avoid an over-interpretation of the data.

For the further discussion it is important to consider the
redistribution of the spectral weight between these spectral
features. For the electronic bands at 1000 and 2500~\cm, the
spectral weight is obtained from the respective Lorentz fit. Since
the Drude behavior does not perfectly describe the zero-frequency
peak in the optical conductivity (especially for the Rb salt), we
determined the respective spectral weight by subtracting the two
Lorentz oscillators from the experimental spectra.

The spectral weight is received by integration of the optical spectra
\cite{DresselGruner02}
\be
\omega^2_p =
8\int_0^{\omega_c}\sigma_1(\omega)\, {\rm d}\omega
\label{eq:sumrule}\quad ;
\ee
the choice of the upper limit $\omega_c$ determines which
excitations are considered. The used cut-off ($\omega_c$) value of
7000~\cm\ is typical for compounds with similar
parameters\cite{Dressel04} because is lies well above the plasma
edge. According to $\omega_{p} = ({4\pi Ne^2}/{m_b})^{1/2}$, where
$N$ is a number of charge carriers known from structural
data~\cite{Mori90}, the spectral weight then yields the band mass
$m_b$. From the spectral weight of the Drude-like contribution
$\left(\omega_p^{\rm Drude}\right)^2$ (see insert (d) in
Fig.~\ref{fig:NH}) we can estimate the effective mass of the
quasi-free charge carriers $m^*$. In the following analysis we
normally use an effective mass of the charge carriers with respect
to the band mass, i.e.\ the ratio $m^*/m_b$. The advantage of this
approach is that then the effective mass value is normalized
according to the thermal contraction of the unit cell.

Within the one dimensional tight-binding approximation, the
spectral weight is related to the width of the bands or the
transfer integral $t$:
\be
\omega_p^2=\frac{16td^2e^2}{\hbar^2V_m}\sin\left\{\frac{\pi}{2}\rho\right\}
\label{eq:plasmaparallel} \quad ; \ee
where $d$ is the inter-molecular distance, $V_m$ denotes the
volume per molecule, and the number of electrons per site is given
by $\rho$. The enlargement of the spectral weight is related to
the increase of the transfer integral; for instance when the
temperature decreases and thus the crystal contracts.\cite{Ono97}
It is commonly believed \cite{Mila,DresselGruner02,Dressel04} that
the total spectral weight is conserved, when the integration in
Eq.~(\ref{eq:sumrule}) is extended to high enough values (above
10\,000~\cm). This is not strictly correct\cite{Drichko06} when
the thermal expansion is extremely large, as in the case of
organic materials where it can be as large as 10 \% \cite{Endo97},
or when external pressure is applied.

\subsection{Assignment of the electronic spectrum}

At first we want to make an assignment of the major spectral
features which are common to the studied BEDT-TTF salts.  The
overdamped maximum around 2500 \cm\ is observed in the spectra of
BEDT-TTF-based organic conductors  with different band
structures;\cite{Dressel04} however, the previously
suggested\cite{Dressel92} assignment to the interband transition
is doubtful. Instead we follow the interpretation\cite{Tajima00}
of the maxima at 1000~\cm\ and 2500~\cm\ to the spectral weight
which is shifted from a zero-peak to higher frequencies due to
influence of electronic correlations.\cite{Merino03}

The calculations of Merino {\it et al.}\cite{Merino03} for
metallic quarter-filled systems close to charge order suggest the
appearance of two maxima in addition to a Drude peak: a sharper
one at $\hbar\omega=2t$ due to short-range charge order, and a
wider one at about $5t$ due to a shift of the spectral weight to
high frequencies on the strong influence of correlations. In a
very good agreement with the experiment, it gives values of 960
and 2500~\cm\ for positions of these features, estimated using $t
= 0.06$~eV, a typical value of transfer integral for these
compounds.  At room temperature in the polarizations parallel and
perpendicular to the stacks the intensity of the two maxima scales
with the plasma frequency of the Drude-like contribution, being
approximately 1.3 times higher for $E\perp$~stacks; at low
temperatures a redistribution of the spectral weight occurs. This
encourages the application of the one-band approximation, used by
theory, for the analysis of our data: i.e.\ all features originate
from one band modulated by electronic correlations.

Interestingly to note, that while the redistribution of the
spectral weight between the features on temperature decrease
(which we discuss later) differs between the compounds, the
frequencies of the electronic bands exhibit a similar temperature
dependence. The low-frequency maximum occurs at the same position
in both polarizations and moves down from 1000~\cm\ to 800~\cm\
when the crystals are cooled to $T=6$~K. The changes in the
position of the higher-frequency maximum are too small compared to
its width and cannot be analyzed.

In order to further prove our interpretation of the electronic
spectra let us follow their change upon reducing the $V/t$
ratio.\cite{Merino03} The effective inter-site interaction $V/t$
can be tuned by external pressure: if the distances between the
molecules decrease, then the overlap integrals will increase. In
fact, the $V$ value will increase as well, but the simple
consideration shows that Coloumb repulsion between electrons on
the neighboring sites $V$ will increase slower, as $1/d$ (where
$d$ is a distance between the molecules), while $t$ will increase
exponentially. Alternatively the same effect is achieved by
thermal contraction of the crystal, even if accompanied by other
temperature-dependent effects. The transfer integrals $t$ of
$\alpha$-(BEDT-TTF)$_2$KHg(SCN)$_4$ calculated for the application
of hydrostatic pressure up to 10~kbar,\cite{Campos96} as well as
those for the K, NH$_4$ and Rb salts calculated from the X-ray
data for a wide temperature range\cite{Ono97} show the same
tendency. Only one of the intra-stack transfer integrals  is
expected to increase on cooling \cite{Ono97}, while the
inter-stack transfer integrals increase by up to 10\%\ on cooling
down to 4~K and up to 25\%\ on the application of pressure up to
10 kbar.

\subsection{Pressure dependence of electronic spectra}

\begin{figure}
\includegraphics[width=7cm]{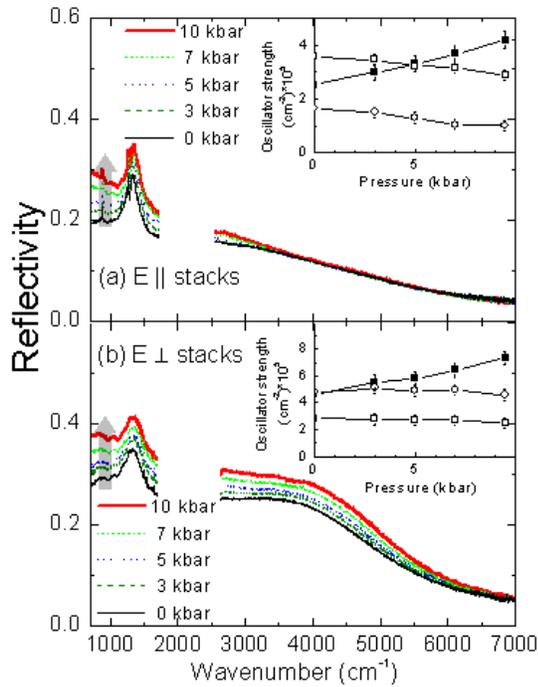}
\caption{\label{fig:Pressure} Color online. Reflectivity and
conductivity spectra of $\alpha$-(BEDT-TTF)$_2$KHg(SCN)$_4$ at 0,
3, 5, 7 and 10 kbar. left panels (a) and (b): $E\parallel$~stacks,
right panels (c) and (d): $E\perp$~stacks. The spectral region
between 1700 and 2500 \cm is cut out from the spectra since it is
affected by the strong absorbance of the diamond cell window.
Arrows indicate the changes with increasing pressure. The inserts
show the pressure dependence of the spectral weight of the Drude
contribution (solid squares), the 1000-\cm\ maximum (open
squares), and the mid-infrared band (open circles).}
\end{figure}

To support our assignment of electronic spectra of the
$\alpha$-(BEDT-TTF)$_2M$Hg(SCN)$_4$ compounds we performed
reflectivity measurements of $\alpha$-(BEDT-TTF)$_2$KHg(SCN)$_4$
under hydrostatic pressure at room temperature in the
600-7000~\cm\ range.  The application of the hydrostatic pressure
modifies the $V/t$ parameter without changing the
temperature-dependent parameters, e.g. scattering rate.
Fig.~\ref{fig:Pressure} shows, that the main tendency in the
spectra on the increasing pressure up to 9.5 kbar is an increase
of reflectivity, for lower frequency region in $E\parallel$~stacks
direction, and in a wider frequency range for $E\bot$~stacks. The
spectral weight of the features estimated by a Drude-Lorentz fit
as proposed in Sec. II, illustrates the main tendency (inserts in
Fig.~\ref{fig:Pressure}): the Drude spectral weight increases on
the expense of the 1000 and 2500~\cm\ maxima. Since the
measurements are done in a comparatively narrow spectral range, we
cannot comment on a change of the total spectral weight. The
reflectivity increase is  more pronounced in $E\bot$~stacks
direction, but in general the changes are not so anisotropic as
was suggested by the transfer integrals
calculations.\cite{Campos96}

The shift of the spectral weight from the high-frequency features
to the Drude-part is expected since the correlations to bandwidth
ratio decreases on rising the transfer integrals: the correlation
effects diminish and the systems becomes more metallic. This
observation strongly supports our conclusion that the mid-infrared
features in the spectra are due to electronic effects. A change of
dimerization is not predicted by the calculations;\cite{Campos96}
in agreement with this no changes are observed in the BEDT-TTF
vibrational features.

\subsection{Temperature dependence of electronic spectra}
\label{sec:mass}

Upon cooling, the electronic properties of these systems are
changed in several respects: (i)~As for any metal, the
charge-carrier scattering rate is reduced since phonons freeze
out. (ii)~The shrinking of the unit-cell volume increases the
number of carriers proportionally. (iii)~The transfer integrals
and bandwidth become larger which reduces the $V/t$ ratio.

\begin{figure}
\includegraphics[width=8.7cm]{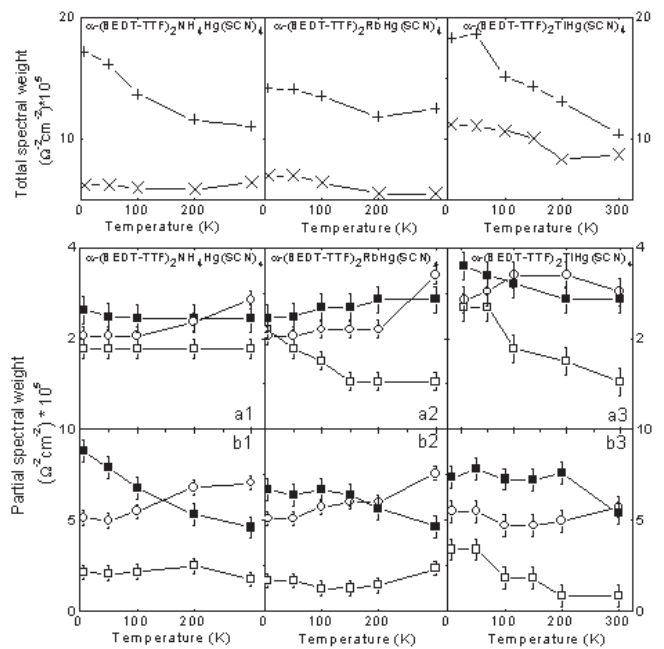}
\caption{\label{spectralweight}Upper panel: temperature dependence
of the total spectral weight for
$\alpha$-(BEDT-TTF)$_2M$Hg(SCN)$_4$ ($M$ = NH$_4$, Rb, Tl)
parallel to the stacks (diagonal crosses) and perpendicular to the
stacks (straight crosses). Lower panel: Temperature dependence of
the spectral weight for the three main features in the optical
spectra of $\alpha$-(BEDT-TTF)$_2M$Hg(SCN)$_4$ ($M$ = NH$_4$, Rb,
Tl) compounds: filled squares - Drude peak, empty squares -
1000~\cm\ maximum, empty circles - 2500~\cm\ maximum. (a)
Polarization parallel to the stacks direction and (b)
perpendicular to the stacks. Note different scales for the
different contributions and polarizations.}
\end{figure}

The total spectral weight increases on cooling due to the thermal
contraction of the unit cell volume (Fig.~\ref{spectralweight})
and an increase of the bandwidth (See Sec.~\ref{DrudeLorentz}). However, in
all cases the variation is more intense in the direction
$E\bot$~stacks in agreement with the anisotropic temperature
behavior calculated for the hopping integrals $t$.\cite{Ono97} The
experimentally observed anisotropy is largest for the NH$_4$
compound, and decreases on going to Rb and Tl.

If no change of the ground state occurs, an increase of the
transfer integrals in the $E\bot$~stacks direction on cooling will
reduce the $V/t$ ratio
 and lead to a redistribution of the spectral weight to the low-frequency features.\cite{Merino03}
Although it is know that $t$ changes as much as 10\%, a
quantitative prediction of the spectral weight shift turns out to
be difficult; experimentally the assessment is easier. Indeed, in
the conductivity spectra [$E\perp$~stacks,
Figs.~\ref{fig:NH}, \ref{fig:Rb}, and \ref{spectralweight}(b)] of the NH$_4$ and Rb
compounds the zero-frequency peak grows on the expense of the
band around 2500~\cm; the tendency is more pronounced for the
NH$_4$ salt.  Interestingly, the redistribution of spectral weight
does basically not involve the low-frequency maximum at 1000~\cm:
the intensity of this feature is conserved. This strengthening  of
the metallic behavior leads to decrease of the effective mass upon
cooling in the NH$_4$ and Rb spectra (Fig.~\ref{MassAndScat}). The
Tl compound, however, shows a different tendency: the intensity of
the maximum at 1000~\cm\ increases on cooling, which is suggestive
for non-metallic behavior.

The same considerations suggest no redistribution of the spectral
weight in the direction parallel to the stacks because the transfer integrals $t$ do not increase upon thermal contraction, that is
confirmed by the constant spectral weight for NH$_4$ and Rb
compound in this direction.

\subsection{In-plane anisotropy} \label{sec:abnisotropy}

One of the unusual features of the studied $\alpha$-phase BEDT-TTF
salts is the in-plane anisotropy. These compounds are considered
to be quasi-two-dimensional conductors, and indeed, the in-plane
conductivity has a metallic character in both directions. With
optical means we can detect an in-plane anisotropy of
approximately a factor of two: perpendicular to the stacks the
conductivity is higher compared to the in-stack direction. This
experimentally detected anisotropy in general agrees with the
calculation of transfer integrals,\cite{Mori_Term} though the
$t$-values are even more anisotropic.

The interesting point is the different temperature behavior for
polarizations $E\parallel$~stacks and $E\perp$~stacks; again the
latter one is `more metallic'.  This is in agreement with the
anisotropy obtained in dc resistivity measurement for
$\alpha$-(BEDT-TTF)$_2$TlHg(SCN)$_4$:\cite{Schegolev} while for
$E\bot$~stacks resistivity is metallic, for the $E\parallel$~stacks
 it shows only a slow decrease with reduced
temperature down to $200-150$~K, while for lower temperatures the
slope becomes steeper.

Maesato {\it et al.}\cite{Maesato} investigated the resistivity
for K and NH$_4$ compounds while applying uniaxial strain. They
proposed that the anisotropy $\rho_c/\rho_a$ defines the ground
state of these salts, being larger for the NH$_4$ superconducting
compound than for the K analog. These observations on in-plane
anisotropy agree very well with our optical measurements which
probe the transfer integrals in the plane: the optical anisotropy
and the anisotropic temperature behavior of the spectral weight
(see Fig.~\ref{spectralweight}) is more pronounced for
$\alpha$-(BEDT-TTF)$_2$NH$_4$Hg(SCN)$_4$ compared to other members
of the family.

\section{Discussion:
charge order fluctuations  $vs.$ static order}

\subsection{Charge fluctuations effects}

\begin{figure}
\includegraphics[width=8.7cm]{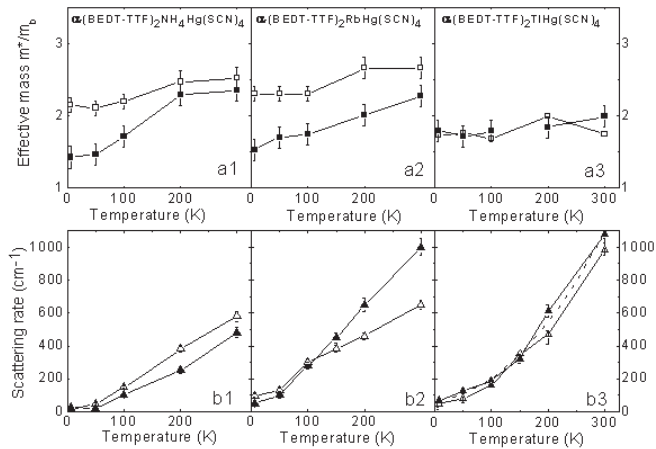}
\caption{\label{MassAndScat}Temperature dependence of the m*/m$_b$
(upper panel) and Drude-scattering rate $\gamma$ (lower panel)
 for $\alpha$-(BEDT-TTF)$_2M$Hg(SCN)$_4$ ($M$ = NH$_4$, Rb, Tl)
compounds: filled squares - $E\bot$~stacks, empty squares
$E\parallel$~stacks. For the Tl-salt a $\gamma \sim T^2$ fit of
the scattering rate is shown with a bold line.}
\end{figure}

The NH$_4$ and Rb compounds show a striking decrease of effective
mass  m*/m$_b$ upon cooling (see Fig.~\ref{MassAndScat}),
especially important is that this effect is seen for
$E\parallel$~stacks polarization. While the strength of the effect
in $E\bot$~stacks direction can be partly explained by the
redistribution of the spectral weight to low frequencies due to
thermal contraction of the crystal (see the above Section and
Fig.~\ref{spectralweight}), this is not the case for
E$\parallel$stacks, where the effective Coulomb repulsion $V/t$
does not change on cooling.\cite{Ono97} Nevertheless, lowering the
temperature from 300~K down to 6~K leads to a reduction of the
effective carrier mass by about 10\% in the direction parallel to
the stacks; the uncertainty in the data and analysis does not
permit to give a functional dependence. Fig.~\ref{spectralweight}
reveals that this effect is caused by the redistribution of the
spectral weight from the mid-infrared to the zero-frequency peak,
again pointing on a decrease of correlation effects.

The observed temperature dependence does not correspond to the
behavior known from simple metals, for which only the scattering
rate increases with temperature, whereas the concentration and
mass of the carriers remain constant. On one hand, this tendency
is opposite to the effects observed in the prime example of
strongly correlated electron systems:\cite{Stewart84,Grewe91} in
heavy fermions the electronic interactions become more important
towards low temperatures and consequently the effective mass is
significantly enhanced by up to three orders of magnitude as
$T\rightarrow 0$. On the other hand, exactly this behavior is
predicted for strongly correlated two-dimensional quarter-filled
metals close to the charge-order phase
transition.\cite{Merino03,Dressel03} Theoretical investigations
indicate that the critical ratio $(V/t)_c$, which separates the
metal from the charge-ordered phase, shifts to the larger values,
as depicted in Fig.~\ref{phasediagram}. When cooling down
vertically (constant $V/t$), we depart from the phase boundary and
hence the system becomes more metallic leading to a decrease of
the effective mass with temperature.\cite{Merino06} The
observations for the NH$_4$ and Rb compounds are well interpreted
as the retreat of these systems from the critical value of
correlations $(V/t)_c$ that corresponds to a reduction of the
effective carrier mass as calculated from the experimental Drude
spectral weight.

The other parameter of the charge carriers, the scattering rate $1/\tau$,
also shows a characteristic temperature behavior. As
demonstrated in Fig.~\ref{MassAndScat}, for the NH$_4$ and Rb
salts the scattering rate linearly changes with temperature down
to a crossover temperature $T^*$ at 50~K; below $T^*$ the decrease
becomes slower.

Charge fluctuations cause a linear temperature dependence
$1/\tau(T) \propto T$ for $T>T^*$,\cite{Merino03,Merino06} as
commonly observed in the case of electrons interacting with
boson-like excitation such as phonons. For $T<T^*$ the scattering
rate increases quadratically, as expected from Fermi-liquid
theory. Thus $T^*$ identifies the temperature where charge-order
fluctuations become important in the metallic state. In the phase
diagram it defines the distance from the critical point at which
the charge ordering transition occurs: $T^* \rightarrow 0$ as $V
\rightarrow V_c$. The characteristic energy scale $k_BT^*$ is very
small, $T^*\ll T_F$, when the system is sufficiently close to the
charge-ordering transition. From the crossover temperature for
NH$_4$, $T^*\approx 50$~K, the transfer integrals can be estimated
to about be approximately 0.06~eV, which is in good agreement with
band structure calculations of these compounds\cite{Mori_Term,
Seo2004} and with the position of the electronic bands in the
spectra.

The observed decrease of the effective mass on cooling and the
linear dependence of the scattering rate on temperature
demonstrate that these quarter-filled systems are close to charge
order and show this particular behavior due to charge-order
fluctuations, while also moving away from the phase border with
charge order on cooling down.

A slope of the temperature dependence of the scattering rate is
strikingly enhanced when going from the NH$_4$ to the Rb compound.
Already in the raw data large differences in the scattering rate
of the Drude-contribution are seen for the various
$\alpha$-(BEDT-TTF)$_2M$Hg(SCN)$_4$ salts: For the NH$_4$ compound
the Drude-like peak becomes very narrow at low temperatures with
an extremely small scattering rate, while for the  Rb salt the
zero-frequency contribution is much wider. Since these compounds
are isostructural, the conduction mechanism should essentially be
the same. One might argue that they contain different impurity and
defect concentrations; however, this would lead to a temperature
independent offset in the scattering rate and cannot explain the
distinct temperature dependence of the scattering rates observed
for these compounds. We suggest, that the increase of the slope on
going from NH$_4$ to Rb compound shows,  the charge fluctuations
are stronger in the latter one.\cite{Merino03}

Indeed, the values of the effective mass are also slightly higher
for the Rb compound. In addition, the less metallic
$E\parallel$~stacks polarization in the Rb salt shows a slight
increase (up to 10\% of intensity) of the 1000~\cm\ feature on
cooling, also suggesting that the system is closer to the border
with charge order.

\subsection{Signatures of ordering in
$\alpha$-(BEDT-TTF)$_2$TlHg(SCN)$_4$}

\begin{figure}
\includegraphics[width=8cm]{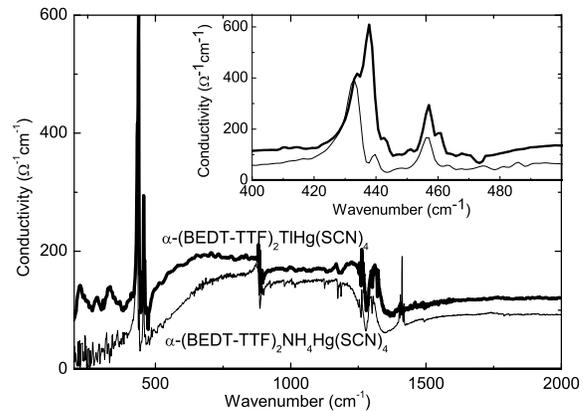}
\caption{\label{vibrations}The T=6 K spectra of
$\alpha$-(BEDT-TTF)$_2$TlHg(SCN)$_4$ (thick line)
$\alpha$-(BEDT-TTF)$_2$NH$_4$Hg(SCN)$_4$ (thin line) in
E$\parallel$~stacks polarization. The insert shows the spectral
range with $\nu_{12}({\rm A}_g)$ and $\nu_{13}({\rm A}_g)$
emv-coupled features for these two spectra. For
$\alpha$-(BEDT-TTF)$_2$TlHg(SCN)$_4$ to electronic band at
1000~\cm\ has higher intensity, and the vibrational features ar
split, pointing on an ordered state.}
\end{figure}
The most important tendency in the temperature behavior of the Tl
compound  is an increase of intensity of the 1000~\cm\ peak on
cooling (see Fig.~\ref{spectralweight}). This feature gets
stronger at $T<200$~K, while the Drude weight and the intensity of
the high-frequency oscillator do not show such a big change.

Above we have assigned the peak around 1000~\cm\ to the
short-range charge-order fluctuations, as suggested by exact
diagonalization calculations .\cite{Merino03} Finding it in the
spectra of all the compounds at room temperature infers that there
is already some charge order present. However, it develops
considerably when the Tl compound is cooled, while the effective
mass in both polarization stays unchanged. This behavior evidences
that in $\alpha$-(BEDT-TTF)$_2$TlHg(SCN)$_4$ short-range charge
order prevails although the compound does not become insulating. A
similar behavior was found for $\alpha$-(BEDT-TTF)$_2$KHg(SCN)$_4$
which remains metallic at any temperature, but shows strong
indications of charge order in the optical
spectra.\cite{Dressel03}

We conclude that the decrease of the effective mass in the NH$_4$
and Rb compounds, on the one hand, and the increase of the
1000~\cm\ feature, on the hand, are competing processes: the two
compounds are separated by a phase boundary or cross-over regime.
Interestingly, for the direction parallel to the stacks the Rb
compound shows a competing behavior below 100 K, inferring the
presence of both effects; thus
$\alpha$-(BEDT-TTF)$_2$RbHg(SCN)$_4$ is located right on the
boundary.

At the moment it is difficult to decide whether stable or
fluctuating charge order dominates in
$\alpha$-(BEDT-TTF)$_2$TlHg(SCN)$_4$. From the temperature
behavior of the scattering rate and vibrational features, we are
inclined to assume the charge order to be more static.  As
indicated by a dashed line in Fig.\ref{MassAndScat}b3, the
scattering rate has a temperature dependence close to a
$T^2$-behavior in the entire temperature range. The origin of such
a $T^2$ dependence in the full temperature range in contrast to
the linear-T dependence of the other salts deserves further
theoretical analysis and seems correlated to the observation of
charge ordering phenomena

In addition to the optical response of the electronic system, we
find some evidence of ordering in the vibrational features of the
Tl compound. Due to the weak screening by the electronic
background, the mode at about 400 \cm\ is clearly seen for the
polarization $E\parallel$~stacks and allows detailed analysis. As
Fig.\ref{vibrations} shows, the $\nu_{12}({\rm A}_g)$ (based on
$D_{2}$ symmetry) vibration shows up as a single band at 457~\cm\
for the NH$_4$ compound, while for
$\alpha$-(BEDT-TTF)$_2$TlHg(SCN)$_4$ the mode is split in two
which are located at 457 and 460~\cm. The band of the
$\nu_{13}({\rm A}_g)$ vibration also splits: three bands at 433,
438, and 443~\cm\ (the latter one is very weak) can be
distinguished for the Tl compound; for the NH$_4$ salt the band at
433~\cm\ is strong while only one weak peak is observed at
440~\cm. Some splitting also observed in the $\nu_6({\rm A}_g)$
anti-resonance feature in the spectrum of the Tl-salt. The Rb salt
is again in the intermediate situation, showing a weak splitting
of the $\nu_{13}({\rm A}_g)$. It should be noted, that the
different splitting of the vibrational features is observed at
temperatures below 200 K, being thus an evidence of a charge
redistribution at the studied temperatures and is not connected to
the low-temperature density wave state; however at higher
temperatures the splitting could not be resolved due to higher
bandwidth.

A closer inspection of the C-N stretching vibrations turns out to
be helpful in our systems. According to the crystal structure,
there are four SCN groups per unit cell; all of them are nearly
parallel to the conducting layers and approximately perpendicular
to the neighbors. Accordingly we observe up to four separate
absorption bands of C-N vibrations. At $T=6$~K when the bands are
well resolved, the NH$_4$ and Rb compounds show two bands in each
polarization; only the Tl salt shows all four C-N bands. This
implies a lower symmetry or slightly different orientation of the
SCN groups in the latter compound.

The lower symmetry of $\alpha$-(BEDT-TTF)$_2$TlHg(SCN)$_4$ is in
agreement with the above conclusion that this compound is closer
to the charge-ordered state than the other salts and a static
charge order is almost established.

\section{Conclusions}

\begin{figure}
\includegraphics[width=9cm]{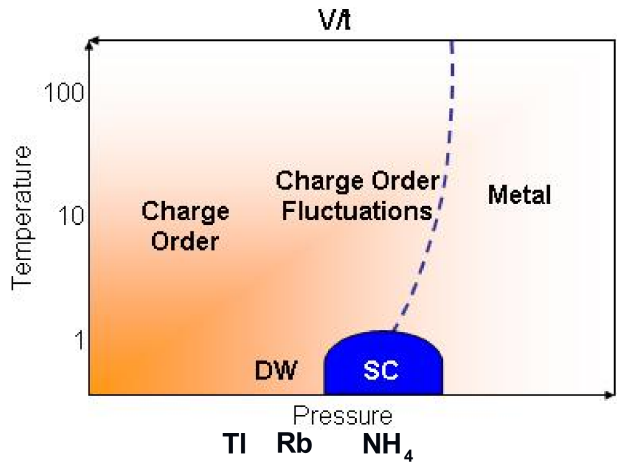}
\caption{\label{phasediagram}Color online. Phase diagram for the
quarter-filled two-dimensional organic conductors
$\alpha$-(BEDT-TTF)$_2M$Hg(SCN)$_4$ ($M$ = NH$_4$, K, Rb, Tl).
With increasing pressure the effective inter-site Coulomb
repulsion $V/t$ is reduced. At intermediate temperatures the
system is tuned from a charge-order state (CO) to a metal with
charge fluctuations in between. We propose that the compounds with
$M$ = K, Tl are to be found in this part of the phase diagram. We
propose that the Rb-salt has lower influence of correlations, and
the NH$_4$ salt is even less correlated, distinctly showing the
increase of metallic properties on cooling, and becoming
superconducting (SC) below 1~K.}
\end{figure}

Our optical studies on the quarter-filled
$\alpha$-(BEDT-TTF)$_2M$Hg(SCN)$_4$ ($M$=NH$_4$, Rb, Tl, K)
evidence that in contrast to the half-filled compounds these
two-dimensional conductors are metals with a Drude peak already
developed at room temperature; but at the same time they show
distinct spectral features caused by charge-order fluctuations.
The spectral weight shift from the zero-frequency conductivity
peak to higher frequencies leads to our assignment of electronic
bands in the spectra. The findings of our optical measurements in
the presence of external pressure confirm this interpretation. The
features due to short-range charge order do not change on
temperature for the NH$_4$ compound, whereas the effective mass of
the charge carriers decreases on cooling. This observation agrees
with the theoretically predicted re-entrance of the charge-order
transition. Stronger electronic correlations in Tl  salt increase
the intensity of the spectral features due to charge order on
cooling; the Rb compound shows the intermediate behavior. From the
slope of the scattering rate and based on comparison to RPA
predictions of a nearly charge ordered two-dimensional metals, we
conclude that correlation effects intensify from NH4 to Rb. In
contrast, Tl displays CO phenomena (as observed in the splitting
of phonon modes) correlated with the presence of a different T²
behavior of scattering rate instead of the linear behavior of NH4
and Rb. This deserves further theoretical and experimental work.
We suggest the order of these materials in a phase diagram in
Fig.~\ref{phasediagram}.

The presence of charge-order fluctuations is observed in all the
studied salts but to a different degree. The less correlated is
the NH$_4$ compound, which shows an increase of metallic
properties on cooling, eventually becomes superconducting below
$T_c=1$~K. Thus we think that it is on the border (marked by a
dashed line) between the charge order fluctuations and metallic
state. On increasing the $V/t$ ratio (moving to the left side of
the phase diagram), this metallic behavior starts to compete with
the charge order: the increase becomes obvious for the Rb salt.
The Tl compound  establishes  short-range charge order upon
cooling, and we would propose that it is the most correlated of
the three studied materials. The same tendency with a strong
increase of the respective spectral feature was observed in the
K-salt.\cite{Dressel03} This is in agreement with the observation
of superconductivity under hydrostatic pressure in
Ref.\onlinecite{Andres2005}

The temperature dependence of the scattering rate supports our
conclusion that the influence of electronic correlations changes
in this range of compounds.

\acknowledgements

We thank M. Calandra, M. Kartsovnik, M. Dumm, R.H. McKenzie, and
H. Seo for helpful discussions. N.D. is grateful for the support
of Alexander von Humboldt Foundation and to  Russian President's
grant "Leading Scientific schools" 5596.2006.2. The project was
partially supported by the Deutsche Forschungsgemeinschaft (DFG),
C.A.K. and A.P. acknowledge a support of Emmy-Noether-program. The
crystal growth at Argonne was performed under the auspices of the
Office of Basic Energy Sciences, Division of Material Sciences of
the U.S. Department of Energy, Contract W-31-109-ENG-38.


\begin{thebibliography}{99}

\bibitem{Mathur}N. D. Mathur, F. M.
Grosche, S. R. Julian, I. R. Walker, D. M. Freye, R. K. W.
Haselwimmer, and G. G. Lonzarich, Nature {\bf 394}, 39 (1998).

\bibitem{Kotliar04}G. Kotliar and D. Vollhardt, Physics Today, March  2004, p.~53.

\bibitem{Seo04}H. Seo, C. Hotta, and H. Fukuyama, Chem. Rev. {\bf 104},
5005 (2004)


\bibitem{McKenzie97}R.H. McKenzie.   {\it Science} {\bf 278}, 821 (1997).

\bibitem{Lefebvre}S. Lefebvre, P. Wzietek, S. Brown, C. Bourbonnais, D. Jerome, C. Meziere, M. Fourmigue, and P. Batail, Phys. Rev. Lett {\bf 85}, 5420 (2000).

\bibitem{Merino01} J. Merino and  R. H. McKenzie, Phys. Rev. Lett. {\bf 87}, 237002 (2001).

\bibitem{Wosnitza96}J. Wosnitza, {\em Fermi Surfaces in Low-Dimensional Organic Metals and Superconductors} (Spinger-Verlag, Berlin, 1996).

\bibitem{Singleton00}J. Singleton, Rep. Prog. Phys. {\bf 63}, 1111 (2000).

\bibitem{Mori_Term}T. Mori, H. Inokuchi, H. Mori, S. Tanaka, M. Oshima, and G. Saito, J. Phys. Soc. Jpn, {\bf 59}, 2624 (1990).

\bibitem{Mori98} H. Mori, S. Tanaka, and T.  Mori, Phys. Rev. B {\bf 57}, 12023 (1998).

\bibitem{Chiba04} R.Chiba, K. Hiraki,  T. Takahashi, H. M.
Yamamoto, T. Nakamura, Phys. Rev. Lett. {\bf 93}, 216405 (2004).

\bibitem{Suzuki03}K. Suzuki, K. Yamamoto,  M. Uruichi,  and
 K. Yakushi, Synth. Met. {\bf 135-136}, 525 (2003).

\bibitem{Schweitzer87} D. Schweitzer, P. Bele, H. Brunner, E. Gogu,
U. Haeberlen, I. Hennig, I. Klutz, R.  Swietlik,  and H.~J.
Keller,  Z. Phys.\ B , {\bf 67},  489
  (1987)

\bibitem{Dressel94} M. Dressel, G. Gr\"uner, J.P. Pouget, A. Breining, and D. Schweitzer, J. Phys. I (France) {\bf 4}, 579 (1994).

\bibitem{Moroto04}S. Moroto, K.-I. Hiraki, Y. Takano, Y. Kubo, T. Takahashi, H. M. Yamamoto, and T. Nakamura. J. Phys. IV (France) {\bf 114} 339 (2004).


\bibitem{DresselGruner02}M.~Dressel and G.~Gr\"{u}ner, {\it Electrodynamics of Solids}  (Cambridge University Press, Cambridge, 2002).
\bibitem{Basov05}D. N. Basov , T. Timusk  Rev. Mod. Phys. {\bf
77}, 721 (2005)

\bibitem{Imada}M. Imada, A. Fujimori, and Y. Tokura
Rev. Mod. Phys. {\bf 70}, 1039 (1998)

\bibitem{Merino03}J. Merino, A. Greco, R. McKenzie, and M. Calandra,

Phys. Rev. B {\bf 68} 245121 (2003).

\bibitem{Dressel03}M. Dressel, N. Drichko, J. Schlueter, and J. Merino,

Phys. Rev. Lett. {\bf 90}, 167002 (2003).

\bibitem{Mori90}H. Mori, S. Tanaka, M. Oshima, G. Saito, T. Mori, Y.
Maruyama, and H. Inokuchi, Bull. Chem. Soc. Jpn., {\bf 63},  2183
(1990).

\bibitem{Homes93}C.C. Homes, M. Reedyk, D.A. Cradles, and T. Timusk, Applied Optics {\bf 32}, 2976 (1993).

\bibitem{Dressel92}M. Dressel, J.E. Eldridge, H.H. Wang, U. Geiser, and J.M. Williams, Synth. Met. {\bf 52}, 201 (1992)

\bibitem{Tajima03}H. Tajima, M. Inoue, R. Sakamoto, J. Yamazaki, and N. Hanasaki, Synth. Met. {\bf 133-134}, 151 (2003).

\bibitem{Mao86}
H. K Mao, J. Xu, and P.M. Bell, J.Geophys. Res [Atmos.]{\bf 91},
4673 (1986).

\bibitem{Kuntscher05}
C. A. Kuntscher, S. Frank, I. Loa, K. Syassen, T. Yamauchi, and Y.
Ueda, Phys. Rev. B {\bf 71}, 220502(R) (2005).


\bibitem{Drichko03}N. Drichko, M. Dressel, A. Kini, and J. Schlueter,

Synth. Met. {\bf 133-134}, 91 (2003).

\bibitem{Kornelsen91}K. Kornelsen, J.E. Eldridge, H.H. Wang, and J.M. Williams, Phys. Rev. B {\bf 44}, 5235 (1991).

\bibitem{Dressel04}M. Dressel and N. Drichko, Chem. Rev. {\bf 104}, 5689 (2004).

\bibitem{Faltermeier05}D. Faltermeier, J. Barz, M. Dumm, N. Drichko, M. Dressel, C. Meziere, and P. Batail, {\it to be published};
M. Dumm, D. Faltermeier, N. Drichko, M. Dressel, C. Meziere, P.
Batail, J. Merino, and R. McKenzie, {\it to be published}
\bibitem{Ono97}S. Ono, T. Mori, S. Endo, N. Toyota, T. Sasaki, Y. Watanabe, and T. Fukase, Physica C {\bf 290}, 49 (1997).
\bibitem{Mila}V. Vescoli, J.Favand, F. Mila, and L. Degiorgi. Eur. Phys. J. B {\bf 3} 149-154
(1998).
\bibitem{Drichko06}N. Drichko and M. Dressel, to be published.
\bibitem{Endo97} S. Endo, W. Watanabe, T. Sasaki, T. Fukase, and
N. Toyota. Synth. Met. {\bf 86}, 2013 (1997)
\bibitem{Tajima00}H. Tajima, S. Kyoden, H. Mori, and S. Tanaka,  Phys. Rev. B {\bf 62}, 9378 (2000).

\bibitem{Campos96}C. Campos,  P. S. Sandhu, J. S. Brooks, T. Ziman, Phys.\ Rev. B {\bf 53}, 12725 (1996).

\bibitem{Schegolev}A.I. Schegolev, V.N. Laukhin, A.G. Khomenko,

M.V. Kartsovnik, R.P. Shibaeva, L.P. Rozenberg, and A.E.

Kovalev, J.\ Phys. I (France) {\bf 2}, 2123 (1992).
 \bibitem{Maesato} M. Maesato, Y. Kaga, R. Kondo, and S.
 Kagoshima. Phys. Rev. B {\bf 64} 155104 (2001)

\bibitem{Stewart84}G.R. Stewart,  Rev.\ Mod.\ Phys.\ {\bf 56}, 755 (1984).
\bibitem{Grewe91}N. Grewe and F. Steglich, in: Handbook on the Physics and
Chemistry  of Rare Earths, Vol.\ {\bf 14}, ed. by K.A. Gscheidner
Jr. and L. Eyring
 (Elsevier, Amsterdam, 1991), p. 343.
\bibitem{Merino06}J. Merino, A. Greco, N. Drichko, M. Dressel. Phys. Rev. Lett. {\bf 96}, 216402 (2006).

\bibitem{Seo2004}
H. Seo, C. Hotta, and H. Fukuyama, Chem. Rev. {\bf 104} (2004)
5005.

\bibitem{Andres2005}D. Andres, M. V. Kartsovnik, W. Biberacher, K.
Neumaier, E. Schuberth, and H. M\"uller. Phys. Rev. B, {\bf 72},
174513 (2005).


\end{thebibliography}
\end{document}